\documentclass[graybox]{svmult}

\usepackage{mathptmx}       % selects Times Roman as basic font
\usepackage{helvet}         % selects Helvetica as sans-serif font
\usepackage{courier}        % selects Courier as typewriter font
\usepackage{type1cm}        % activate if the above 3 fonts are
\usepackage{makeidx}         % allows index generation
\usepackage{graphicx}        % standard LaTeX graphics tool
\usepackage{multicol}        % used for the two-column index
\usepackage[bottom]{footmisc}% places footnotes at page bottom

\usepackage{amsmath}

\makeindex             % used for the subject index
\begin{document}

\title*{Community-Wide Evaluation of Computational Function Prediction}
\author{Iddo Friedberg and Predrag Radivojac}
\institute{Iddo Friedberg \at Department of Veterinary Microbiology and Preventive Medicine, Iowa State University, Ames, IA 50011, USA; \email{idoerg@iastate.edu}
\and Predrag Radivojac \at Department of Computer Science and Informatics, Indiana University, Bloomington, IN 47405, USA; \email{predrag@indiana.edu}}
\maketitle

\abstract{A biological experiment is the most reliable way of assigning function to a protein. However, in the era of high-throughput sequencing, scientists are unable to carry out experiments to determine the function of every single gene product. Therefore, to gain insights into the activity of these molecules and guide experiments, we must rely on computational means to functionally annotate the majority of sequence data. To understand how well these algorithms perform, we have established a challenge involving a broad scientific community in which we evaluate different annotation methods according to their ability to predict the associations between previously unannotated protein sequences and Gene Ontology terms. Here we discuss the rationale, benefits and issues associated with evaluating computational methods in an ongoing community-wide challenge.
}

% \abstract{Each chapter should be preceded by an abstract (10--15 lines long) that summarizes the content. The abstract will appear \textit{online} at \url{www.SpringerLink.com} and be available with unrestricted access. This allows unregistered users to read the abstract as a teaser for the complete chapter. As a general rule the abstracts will not appear in the printed version of your book unless it is the style of your particular book or that of the series to which your book belongs.\newline\indent
% Please use the 'starred' version of the new Springer \texttt{abstract} command for typesetting the text of the online abstracts (cf. source file of this chapter template \texttt{abstract}) and include them with the source files of your manuscript. Use the plain \texttt{abstract} command if the abstract is also to appear in the printed version of the book.}

\section{Introduction}
\label{sec:1}
Molecular biology has become a high-volume information science. This rapid transformation has taken place over the past two decades and has been chiefly enabled by two technological advances: (\emph{i}) affordable and accessible high-throughput sequencing platforms, sequence diagnostic platforms and proteomic platforms, and (\emph{ii}) affordable and accessible computing platforms for managing and analyzing these data. %As a result of these technological advances, as well as the expanding size of the scientific community interested in biomedical science, the volume of molecular data has been rapidly rising. For example, 
It is estimated that sequence data accumulates at the rate of 100 exabases per day (1 exabase~$=10^{18}$ bases) \cite{Stephens2015Big}. However, the available sequence data are of limited use without understanding their biological implications. Therefore, the development of computational methods that provide clues about functional roles of biological macromolecules is of primary importance. 

Many function prediction methods have been developed over the past two decades \cite{Friedberg2006Automated, Rentzsch2009Protein}. Some are based on sequence alignments to proteins for which the function has been experimentally established \cite{Martin2004, Engelhardt2005, Clark2011}, yet others exploit other types of data such as protein structure \cite{Pazos2004, Pal2005}, protein and gene expression data \cite{Huttenhower2006}, macromolecular interactions \cite{Letovsky2003, Nabieva2005}, scientific literature \cite{Camon2005}, or a combination of several data types \cite{Troyanskaya2003, Sokolov2010, Cozzetto2013Protein}.  %\textcolor{red}{P.R. I'd also like to include David Jones' CAFA1 method here - they used multiple data types. Maybe also one of those genomic context methods? Iddo: I added David Jones.}
Typically, each new method is trained and evaluated on different data. Therefore, establishing best practices in method development and evaluating the accuracy of these methods in a standardized and unbiased setting is important. To help choose an appropriate method for a particular task, scientists often form community challenges for evaluating methods \cite{Costello2013}. The scope of these community challenges extends beyond testing methods: they have been successful in invigorating their respective fields of research by building communities and producing  new ideas and collaborations (e.g. \cite{Kryshtafovych2014CASP10}).

In this chapter we discuss a community-wide effort whose goal is to help understand the state of affairs in computational protein function prediction and drive the field forward. We are holding a series of challenges which we named the Critical Assessment of Functional Annotation, or CAFA. CAFA was first held in 2010-2011 (CAFA1) and included 23 groups from 14 countries who entered 54 computational function prediction methods that were assessed for their accuracy. To the best of our knowledge, this was the first large-scale effort to provide insights into the strengths and weaknesses of protein function prediction software in the bioinformatics community. CAFA2 was held in 2013-2014, and more than doubled the number of groups (56) and participating methods (126). Although several repetitions of the CAFA challenge would likely give accurate trajectory of the field, there are valuable lessons already learned from the two CAFA efforts.%Here we describe the reasons that led us to a specific organization of the CAFA experiment and conclusions that we arrived at. 

For further reading on CAFA1, the results were reported in full in \cite{Radivojac2013LargeScale}. As of this time, the results of CAFA2 are still unpublished and will be reported in the near future. The preprint of the paper is available on arXiv \cite{Jiang2015}.

\section{Organization of the CAFA challenge} 
\begin{figure}
\sidecaption
% For example, with the graphicx style use
%\includegraphics[scale=.65]{cafa_timeline.png}
\includegraphics[width=\textwidth]{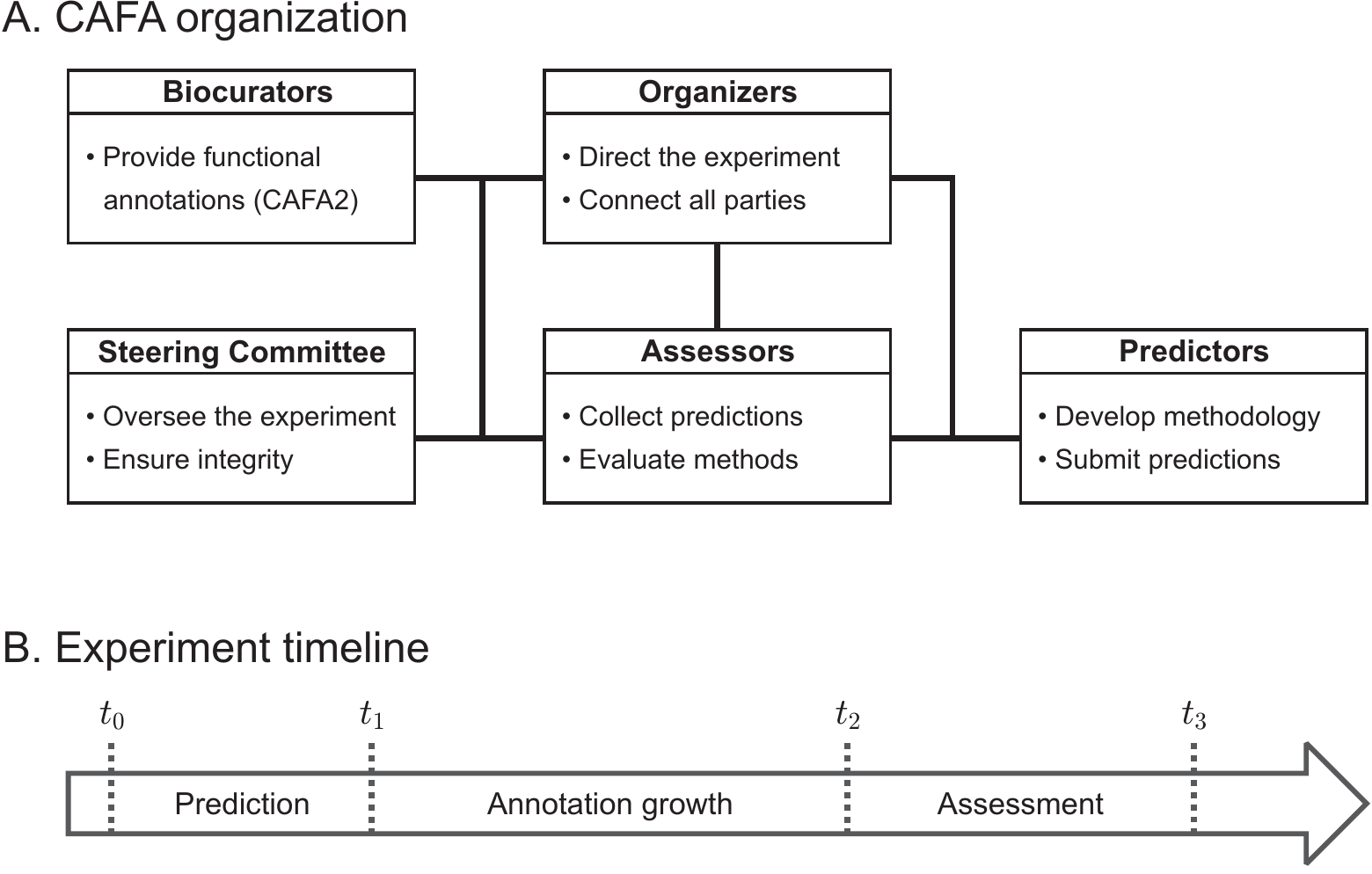}
%
% If no graphics program available, insert a blank space i.e. use
%\picplace{5cm}{2cm} % Give the correct figure height and width in cm
%
\caption{\textbf{The organizational structure of the CAFA experiment.} (A) Five groups of participants in the experiment together with their main roles. Organizers, assessors and biocurators cannot participate as predictors. (B) Timeline of the experiment.}
\label{fig:cafa_time}       % Give a unique label
\end{figure}

We begin our explanation of CAFA by describing the participants. The CAFA challenge generally involves the following groups: the organizers, the assessors, the biocurators, the steering committee, and the predictors (Figure~\ref{fig:cafa_time}A). 

The main role of the organizers is to run CAFA smoothly and efficiently. They advertise the challenge to recruit predictors, coordinate activities with the assessors, report to the steering committee, establish the set of challenges and types of evaluation, and run the CAFA web site and social networks. The organizers also compile CAFA data and coordinate the publication process. %Despite being so terribly busy, the organizers are pretty fucking awesome (no joke, we actually mean it!). 
The assessors develop assessment rules, write and maintain assessment software, collect the submitted prediction data, assess the data, and present the evaluations to the community. The assessors work together with the organizers and the steering committee on standardizing submission formats and developing assessment rules. The biocurators joined the experiment during CAFA2: they provide additional functional annotations that may be particularly interesting for the challenge. The steering committee members are in regular contact with the organizers and assessors. They provide advice and guidance that ensures the quality and integrity of the experiment. Finally, the largest group, the predictors, consists of research groups who develop methods for protein function prediction and submit their predictions for evaluation. The organizers, assessors and biocurators are not allowed to officially evaluate their own methods in CAFA.

CAFA is run as a timed-challenge (Figure~\ref{fig:cafa_time}B). At time $t_0$, a large number of experimentally unannotated proteins are made public by the organizers and the predictors are given several months, until time $t_1$, to upload their predictions to the CAFA server. At time $t_1$ the experiment enters a waiting period of at least several months, during which the experimental annotations are allowed to accumulate in databases such as Swiss-Prot \cite{Bairoch2005} and UniProt-GOA \cite{Huntley2015}. These newly accumulated annotations are collected at time $t_2$ and are expected to provide experimental annotations for a subset of original proteins. The performance of participating methods is then analyzed between time points $t_2$ and $t_3$ and presented to the community at time $t_3$. It is important to mention that unlike some machine learning challenges, CAFA organizers do not provide training data that is required to be used. CAFA, thus, evaluates a combination of biological knowledge, the ability to collect and curate training data and the ability to develop advanced computational methodology.

%provide a portal for uploading predictions from the community,

We have previously described some of the principles that guide us in organizing CAFA \cite{Friedberg2015}. It is important to mention that CAFA is associated with the Automated Function Prediction Special Interest Group (Function-SIG) that is regularly held at the Intelligent Systems for Molecular Biology (ISMB) conference \cite{Wass2014Automated}. These meetings provide a forum for exchanging ideas and communicating research among the participants. Function-SIG also serves as the venue at which CAFA results are initially presented and where the feedback from the community is sought.

% \section{The Gene Ontology graph determines functional repertoire} Iddo: No it doesn't! WTF? The GO *captures our knowledge* of currently known functional repertoire, if anything. It's a sad day for science and philosophy when ontology is confused with epistemology. 
%\section{The Gene Ontology structure reflects the known functional repertoire}
%\section{The Gene Ontology structure establishes the functional}
\section{The Gene Ontology provides the functional repertoire for CAFA}
\label{sec:go-info}
Computational function prediction methods have been reviewed extensively \cite{Friedberg2006Automated, Rentzsch2009Protein} and are also discussed in the chapter by Cozzetto \& Jones. Briefly, a function prediction method can be described as a classifier: an algorithm that is tasked with correctly assigning biological function to a given protein. This task, however, is arbitrarily difficult unless the function comes from a finite, preferably small, set of functional terms. Thus, given an unannotated protein sequence and a set of available functional terms, a predictor is tasked with associating terms to a protein, giving a score (ideally, a probability) to each association.

The Gene Ontology (GO) \cite{Ashburner2000Gene} is a natural choice when looking for a standardized, controlled vocabulary for functional annotation. GO's high adoption rate in the protein annotation community helped ensure CAFA's attractiveness, as many groups were already developing function prediction methods based on GO, or could  migrate their methods to GO as the ontology of choice. A second consideration is GO's ongoing maintenance: GO is continuously maintained by the Gene Ontology Consortium, edited and expanded based on ongoing discoveries related to the function of biological macromolecules.

%Manual function annotation using GO is, at first blush, a straightforward task: knowing the protein's function or functions, the biocurator selects the most appropriate GO terms and associates them with that protein. In addition, the annotator provides one or more evidence codes for each annotation, to justify the decision made in the annotation. However, the devil is in the details, and proper methods and working standards are employed, as well as a good working knowledge of biological science, and the Gene Ontology's structure and ability to capture information.

One useful characteristic of the basic GO is that its directed acyclic graph structure can be used to quantify the information provided by the annotation; for details on the GO structure see the chapter by Munoz-Torres \textit{et al.} Intuitively, this can be explained as follows: the annotation term ``Nucleic acid binding" is less specific than ``DNA binding" and, therefore, is less informative (or has a lower \textit{information content}). (A more precise definition of information content and its use in GO can be found in \cite{Lord2003Semantic, Schnoes2013Biases}.) The following question arises: if we know that the protein is annotated with the term ``Nucleic acid binding", how can we quantify the additional information provided by the term ``DNA binding" or incorrect information provided by the term ``RNA binding"? The hierarchical nature of GO is therefore important in determining proper metrics for annotation accuracy. The way this is done will be discussed in Section \ref{sec:assessing_quality}.

When annotating a protein with one or more GO terms, the association of each GO term with the protein should be described using an Evidence Code (EC), indicating how the annotation is supported. For example, the Experimental Evidence code (EXP) is used in an annotation to indicate that an experimental assay has been located in the literature, whose results indicate a gene product's function. Other experimental evidence codes include Inferred by Expression Pattern (IEP), Inferred from Genetic Interaction (IGI), and Inferred from Direct Assay (IDA), among others. Computational evidence codes include lines of evidence that were generated by computational analysis, such as orthology (ISO), genomic context (IGC), or identification of key residues (IKR). Evidence codes are not intended to be a measure of trust in the annotation, but rather a measure of provenance for the annotation itself. However, annotations with experimental evidence are regarded as more reliable than computational ones, having a provenance stemming from experimental verification. In CAFA, we treat proteins annotated with experimental evidence codes as a ``gold standard'' for the purpose of assessing predictions, as explained in the next section. The computational evidence codes are treated as predictions.
%Need to shorten, just introduce the evidence codes.

% Iddo: confusing here. We treat nonexperimental evidence codes simply as predictions.
%\textcolor{red}{P.R. Maybe we could mention here that GO Consortium will be transitioning to the evidence code ontology. Raechel Huntley has a paper on this I think last year - with other folks...Iddo: I think that's a distraction. We would need to explain why ECO, difference between ECO and EC, etc. etc. Not the place, IMHO}

%From the point of view of a computational challenge, it is important to emphasize its size and hierarchical structure because both significantly influence method development and evaluation. 
%In particular, GO can be seen as a directed acyclic graph where the nodes correspond to functional terms and edges correspond to the relational ties between pairs of functional terms. The relational ties are directional in that they readily establish parent-child relationships in the ontology; e.g., $\mathsf{is}$-$\mathsf{a}$ or $\mathsf{part}$-$\mathsf{of}$ edges. 
From the point of view of a computational challenge, it is important to emphasize that the hierarchical nature of the GO graph leads to the property of \emph{consistency} in functional annotation. Consistency means that when annotating a protein with a given GO term, it is automatically annotated with all the ancestors of that term. For example, a valid prediction cannot include ``DNA binding" but exclude ``Nucleic acid binding" from the ontology because DNA binding implies nucleic acid binding. We say that a prediction is not consistent if it includes a child term, but excludes its parent. In fact, the UniProt resource and other databases do not even list these parent terms from a protein's experimental annotation. If a protein is annotated with several terms, a valid complete annotation will automatically include all parent terms of the given terms,  propagated to the root(s) of the ontology. The result is that a protein's annotation can be seen as a consistent sub-graph of GO. Since any computational method effectively chooses one of a vast number of possible consistent sub-graphs as its prediction, the sheer size of the functional repertoire suggests that function prediction is non-trivial.

\section{Comparing the performance of prediction methods}
\label{sec:performance}
In the CAFA challenge, we ask the participants to associate a large number of proteins with GO terms and provide a probability score for each such association. Having associated a set of GO sub-graphs with a given confidence, the next step is to assess how accurate these predictions are. This involves: ($i$) establishing standards of truth and ($ii$) establishing a set of assessment metrics.

\subsection{Establishing standards of truth}
The main challenge to establishing a standard-of-truth set for testing function prediction methods is to find a large set of correctly annotated proteins whose functions were, until recently, unknown. An obvious choice would be to ask experimental scientists to provide these data from their labs. However, scientists prefer to keep the time between discovery and publication as brief as possible, which means that there is only a small window in which new experimental annotations are not widely known and can be used for assessment. Furthermore, each experimental group has its own ``data sequestration window'' making it hard to establish a common time for all data providers to sequester their data. Finally, to establish a good statistical baseline for assessing prediction method performance, a large number of prediction targets are needed, which is problematic since most laboratories research one or only a few proteins each. High-throughput experiments, on the other hand, provide a large number of annotations, but those tend to concentrate only on few functions, and generally provide annotations that have a lower information content  \cite{Schnoes2013Biases}.

Given these constraints, we decided that CAFA would not initially rely on direct communication between the CAFA organizers and experimental scientists to provide new functional data. Instead, CAFA relies primarily on established biocuration activities around the world: we use annotation databases to conduct CAFA as a time-based challenge. To do so, we exploit the following dynamics that occurs in annotation databases: protein annotation databases grow over time. Many proteins that at a given time $t_1$ do not have experimentally-verified annotation, but later, some of proteins may gain experimental annotations, as biocurators add these data into the databases. This subset of proteins that were not experimentally annotated at $t_1$, but gained experimental annotations at $t_2$, are the ones that we use as a test set during assessment (Figure~\ref{fig:cafa_time}B). In CAFA1 we reviewed the growth of Swiss-Prot over time and chose 50,000 \textit{target proteins} that had no experimental annotation in the Molecular Function or Biological Process ontologies of GO. At $t_2$, out of those 50,000 targets we identified 866 \textit{benchmark proteins}; i.e., targets that gained experimental annotation in the Molecular Function and/or Biological Process ontologies. While a benchmark set of 866 proteins constitutes only 1.7\% of the number of original targets, it is large enough set for assessing performance of prediction methods. To conclude, exploiting the history of the Swiss-Prot database enabled its use as the source for standard-of-truth data for CAFA. In CAFA2, we have also considered experimental annotations from UniProt-GOA \cite{Huntley2015} and established 3,681 benchmark proteins out of 100,000 targets (3.7\%).

One criticism of a time-based challenge is that when assessing predictions, we still may not have a full knowledge of a protein's function. A protein may have gained experimental validation for function $f_1$, but it may also have another function, say $f_2$, associated with it, which has not been experimentally validated by the time $t_2$. A method predicting $f_2$ may be judged to have made a false-positive prediction, even though it is correct (only we do not know it yet). This problem, known as the ``incomplete knowledge problem'' or the ``open world problem'' \cite{Dessimoz2013}  
is discussed in detail in the chapter by Skunca \textit{et al}. Although the incomplete knowledge problem may impact the accuracy of time-based evaluations, its actual impact in CAFA has not been substantial. There are several reasons for this, including the robustness of the evaluation metrics used in CAFA, and that the newly added terms may be unexpected and more difficult to predict. The influence of incomplete data and conditions under which it can  affect a time-based challenge were investigated and discussed in \cite{Jiang2014Impact}. Another criticism of CAFA is that the experimental functional annotations are not unbiased because some terms have a much higher frequency than others due to artificial considerations. There are two chief reasons for this bias: first, high-throughput assays typically assign shallow terms to proteins, but being high throughput means they can dominate the experimentally-verified annotations in the databases. Second, biomedical research is driven by specific areas of human health, resulting in over-representation of health-related functions \cite{Schnoes2013Biases}. Unfortunately, CAFA1 and CAFA2 could not guarantee unbiased evaluation. However, we will expand the challenge in CAFA3 to collect genome-wide experimental evidence for several biological terms. Such an assessment will result in unbiased evaluation on those specific terms.
 
\subsection{Assessment metrics}
\label{sec:assessing_quality}
When assessing the prediction quality of different methods, two questions come to mind. First, what makes a good prediction? Second, how can one score and rank prediction methods? There is no simple answer to either of these questions. As GO comprises three ontologies that deal with different aspects of biological function, different methods should be ranked separately with respect to how well they perform in Molecular Function, Biological Process, or the Cellular Component ontologies. Some methods are trained to predict only for a subset of any given GO graph. For example, they may only provide predictions of DNA-binding proteins or of mitochondrial-targeted proteins. Furthermore, some methods are trained only on a single species or a subset of species (say, eukaryotes), or using specific types of data such as protein structure, and it does not make sense to test them on benchmark sets for which they were not trained. To address this issue, CAFA scored methods both in general performance, but also on specific subsets of proteins taken from humans and  model organisms, including \textit{Mus musculus}, \textit{Rattus norvegicus}, \textit{Arabidopsis thaliana}, \textit{Drosophila melanogaster}, \textit{Caenorhabditis elegans}, \textit{Saccharomyces cerevisiae}, \textit{Dictyostelium discoideum}, and \textit{Escherichia coli}. In CAFA2, we extended this evaluation to also assess the methods only on benchmark proteins on which they made predictions; i.e., the methods were not penalized for omitting any benchmark protein.

One way to view function prediction is as an information retrieval problem, where the most relevant functional terms should be correctly retrieved from GO and properly assigned to the amino-acid sequence at hand. Since each term in the ontology implies some or all of its ancestors,\footnote{Some types of edges in Gene Ontology violate the transitivity property (consistency assumption), but they are not frequent.} a function prediction program's task is to assign the best consistent sub-graph of the ontology to each new protein and output a prediction score for this sub-graph and/or each predicted term. An intuitive scoring mechanism for this type of problem is to treat each term independently and provide the precision-recall curve. We chose this evaluation as our main evaluation in CAFA1 and CAFA2. 

\begin{figure}[t]
\sidecaption
% For example, with the graphicx style use
%\includegraphics[scale=.85]{precision_recall.png}
%\includegraphics[width=\textwidth]{pics/metrics.pdf}
\includegraphics[width=\textwidth]{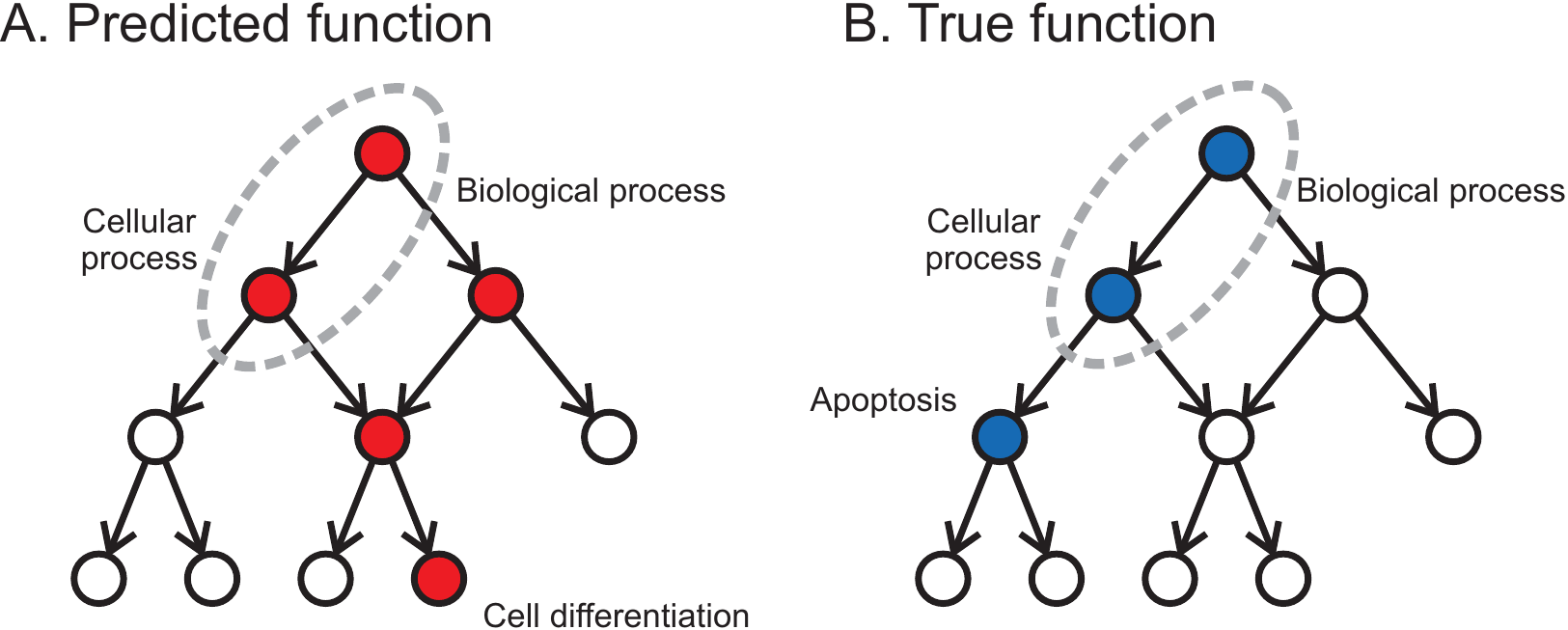}
%
% If no graphics program available, insert a blank space i.e. use
%\picplace{5cm}{2cm} % Give the correct figure height and width in cm
%
\caption{\textbf{CAFA assessment metrics.} (A) Red nodes are the predicted terms $P$ for a particular decision threshold in a hypothetical ontology and (B) blue nodes are the true, experimentally determined terms $T$. The circled terms represent the overlap between the predicted subgraph and the true subgraph. There are two nodes (circled) in the intersection of $P$ and $T$, whereas $|P|=5$ and $|T|=3$. This sets the prediction's precision at $\frac{2}{5}=0.4$ and recall at $\frac{2}{3}=0.667$, with $F_1= 2\times \frac{0.4\times 0.667}{0.4+0.667} = 0.5$. The remaining uncertainty ($ru$) is the information content of the uncircled blue node in panel B, while the misinformation ($mi$) is the total information content of the uncircled red nodes in panel A. An information content of any node $v$ is calculated from a representative database as $-\log \Pr(v|\textrm{Pa}(v))$; i.e., the probability that the node is present in a protein's annotation given that all its parents are also present in its annotation.}
%Would LIKE TO ADD "CELLULAR PROCESS" in second node from top, left side. 
\label{fig:cafa_metrics}       % Give a unique label
\end{figure}

\begin{figure}[t]
\sidecaption
% For example, with the graphicx style use
%\includegraphics[scale=.85]{precision_recall.png}
%\includegraphics[width=\textwidth]{pics/prec-rec-curves.png}
%\includegraphics[width=\textwidth]{pics/curves.pdf}
\includegraphics[width=\textwidth]{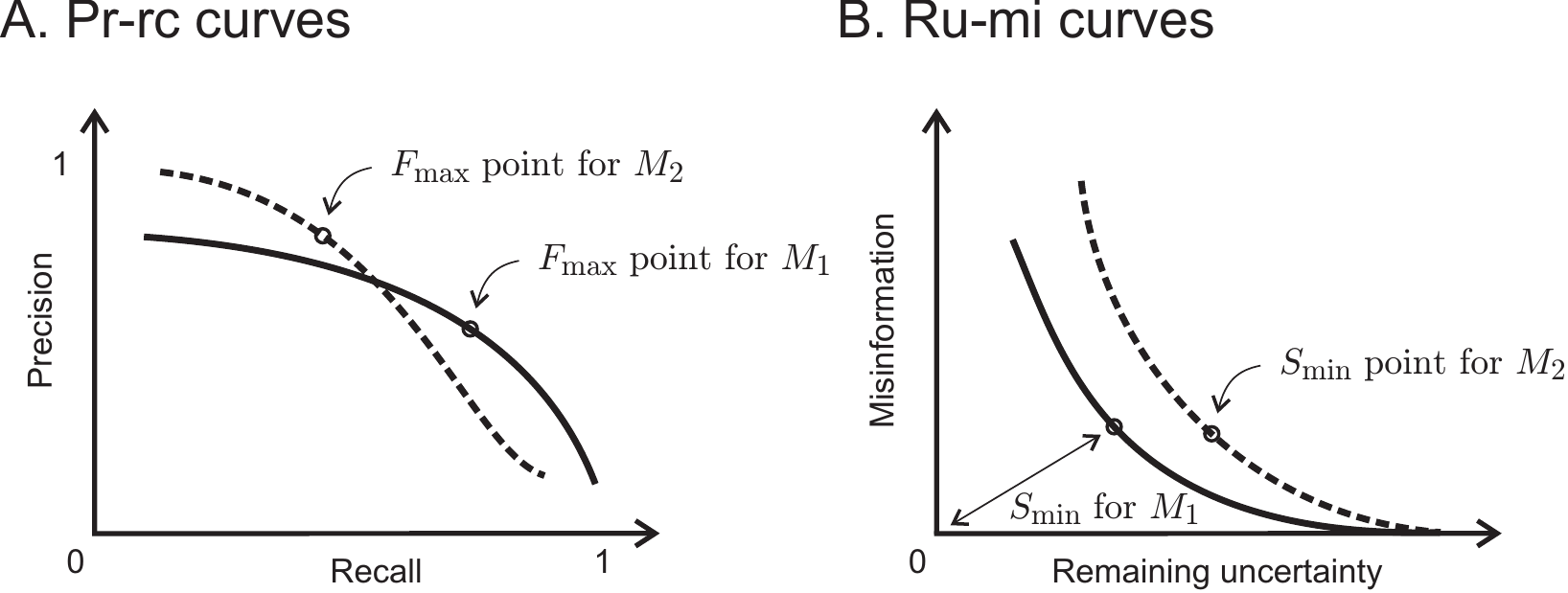}
%
% If no graphics program available, insert a blank space i.e. use
%\picplace{5cm}{2cm} % Give the correct figure height and width in cm
%
\caption{\textbf{Precision-recall curves and remaining uncertainty-misinformation curves.} This figure illustrates the need for multiple assessment metrics, and understanding  the context in which the metrics are used. (A) two pr-rc curves corresponding to two prediction methods $M_1$ and $M_2$. The point on each curve that gives $F_{\max}$ is marked as a circle. Although the two methods have a similar performance according to $F_{\max}$, method $M_1$ achieves its best performance at high recall values, whereas method $M_2$ achieves its best performance at high precision values. (B) two ru-mi curves corresponding to the same two prediction methods  with marked points where the minimum semantic distance is achieved. Although the two methods have similar performance in the pr-rc space, method $M_1$ outperforms $M_2$ in ru-mi space. Note, however, that the performance in ru-mi space depends on the frequencies of occurrence of every term in the database. Thus, two methods may score differently in their $S_{\min}$ when the reference database changes over time, or using a different database.
}

\label{fig:prec-rec-curves}       % Give a unique label
\end{figure}

Let us provide more detail. Consider a single protein on which evaluation is carried out, but keep in mind that CAFA eventually averages all metrics over the set of benchmark proteins. Let now $T$ be a set of experimentally-determined nodes and $P$ a non-empty set of predicted nodes in the ontology for the given protein. Precision ($pr$) and recall ($rc$) are defined as
%$$
%   pr = \frac{tp}{tp+fp}; \quad
%   rc = \frac{tp}{tp+fn},
%$$

$$
pr(P,T)=\frac{|P\cap T|}{|P|}; \qquad rc(P,T)=\frac{|P\cap T|}{|T|},
$$

%\noindent where $tp$ is the number of true positives (experimentally determined terms correctly predicted), $fp$ is the number of false positives and $fn$ is the number of false negatives.  

\noindent where $|P|$ is the number of predicted terms, $|T|$ is the number of experimentally-determined terms, and $|P \cap T|$ is the number of terms appearing in both $P$ and $T$; see Figure~\ref{fig:cafa_metrics} for an illustrative example of this measure. Usually, however, methods will associate scores with each predicted term and then a set of terms $P$ will be established by defining a score threshold $t$; i.e., all predicted terms with scores greater than $t$ will constitute the set $P$. By varying the decision threshold $t \in [0, 1]$, the precision and recall of each method can be plotted as a curve $(pr(t), rc(t))_t$, where one axis is the precision and the other the recall; see Figure~\ref{fig:prec-rec-curves} for an illustration of pr-rc curves and \cite{Radivojac2013LargeScale} for pr-rc curves in CAFA1. To compile the precision-recall information into a single number that would allow easy comparison between methods, we used the maximum harmonic mean of precision and recall anywhere on the curve, or the maximum  $F_1$-measure which we call $F_{\max}$
%$$
%    F_1 = 2\times\frac{pr\times rc}{pr + rc}
%$$
%$$
%     F_{\max} = \max \left \{ 2\times\frac{pr(t)\times rc(t)}{pr(t)+rc(t)} \right \}.
%$$
\[
F_{\max} = \underset{t}{\max}\left \{ 2\times\frac{pr(t)\times rc(t)}{pr(t)+rc(t)} \right \}, 
\]
%On the precision-recall curve, $F_{max}$ is that point that is the closest to the (1,1) coordinate value, and is the best performance of the method over the range of thresholds. 
%\noindent where we modified $pr(t)$ and $rc(t)$ to reflect the dependency on $t$. This measure is preferred because prediction methods are not required to give different scores to different terms in the ontology and may simply choose to predict a small set of terms associating all of them with the same non-zero score. In such cases, $F_{\max}$ can effectively compare methods that assign scores over a range of thresholds with those that do not. In such cases, establishing the area under the precision-recall curve would be meaningless. IDDO: THE AUC was never mentioned before and the reader would be confused as to what an AUC is, and why you are talking about it here. Yes, Christophe did mention that in the review, but unless you want to explain what an AUC is, how prevalent it is, and end up with  why we don't use it, I suggest we just drop this whole thing.

\noindent where we modified $pr(t)$ and $rc(t)$ to reflect the dependency on $t$. It is worth pointing out that the F-measure used in CAFA places equal emphasis on precision and recall as it is unclear which of the two should be weighted more. One alternative to $F_1$ would be the use of a combined measure that weighs precision over recall, which reflects the preference of many biologists for few answers with a high fraction of correctly predicted terms (high precision) over many answers with a lower fraction of correct predictions (high recall); the rationale for this tradeoff is illustrated in Figure~\ref{fig:prec-rec-curves}. However, preferring precision over recall in a hierarchical setting can steer methods to focus on shallow (less informative) terms in the ontology and thus be of limited use. At the same time, putting more emphasis on recall may lead to overprediction, a situation in which many or most of the predicted terms are incorrect. For this reason, we decided to equally weight precision and recall. Additional metrics within the precision-recall framework have been considered, though not implemented yet.  %One alternative would be a general form of the harmonic mean of precision and recall
%$$
%    F_{\beta} = (1+\beta^2)\times \frac{pr\times rc}{(\beta^2\times pr) + rc},
%$$
%\noindent where $\beta$ can be chosen to differentially weight precision and recall.  

%\begin{figure}%[!h]

Precision and recall are  useful because they are easy to interpret: a precision of $\frac{1}{2}$ means that one half of all predicted terms are correct, where a recall of $\frac{1}{3}$ means that a third of the experimental terms have been recovered by the predictor. Unfortunately, precision-recall curves and $F_1$, while simple and interpretable measures for evaluating ontology-based predictions, are limited because they ignore the hierarchical nature of the ontology and dependencies among terms. They also do not directly capture the information content of the predicted terms.

Assessment metrics that take into account the information-content of the terms were developed in the past \cite{Lord2003Semantic, Lord2003Investigating, Pesquita2009Semantic}, and are also detailed in the chapter by Pesquita. In CAFA2 we used an information-theoretic measure in which each term is assigned a probability that is dependent on the probabilities of its direct parents. These probabilities are calculated from the frequencies of the terms in the database used to generate the CAFA targets. The entire ontology graph, thus, can be seen as a simple Bayesian network \cite{Clark2013Informationtheoretic}. Using this representation, two information-theoretic analogs of precision and recall can be constructed. We refer to these quantities as \emph{misinformation} ($mi$), the information content attributed to the nodes in the predicted graph that are incorrect, and \emph{remaining uncertainty} ($ru$), the information content of all nodes that belong to the true annotation but not the predicted annotation. More formally, if $T$ is a set of experimentally-determined nodes and $P$ a set of predicted nodes in the ontology, then
$$
   ru(P, T)=-\sum_{v\in T-P}\log \Pr(v|\textrm{Pa}(v)); \quad
   mi(P, T)=-\sum_{v\in P-T}\log \Pr(v|\textrm{Pa}(v)),
$$
\noindent where $\textrm{Pa}(v)$ is the set of parent terms of the node $v$ in the ontology (Figure~\ref{fig:cafa_metrics}). A single performance measure to rank methods, the minimum semantic distance $S_{\min}$, is the minimum distance from the origin to the curve $(ru(t), mi(t))_t$. It is defined as
%\[
%     S_{\min} = \min(ru^k(t) + mi^k(t))^{\frac{1}{k}}
%\]
\[
S_{\min} = \underset{t}{\min}\left \{ (ru^k(t) + mi^k(t)) ^{\frac{1}{k}} \right \}, 
\]

\noindent where $k \geq 1$. We typically choose $k=2$, in which case $S_{\min}$ is the minimum Euclidean distance between the ru-mi curve and the origin of the coordinate system (Figure~\ref{fig:prec-rec-curves}B). The ru-mi plots and $S_{\min}$ metrics compare the true and predicted annotation graphs by adding an additional weighting component to high-information nodes. In that manner, predictions with a higher information content will be assigned larger weights. The semantic distance has been a useful measure in CAFA2 as it properly accounts for term dependencies in the ontology. However, this approach also has limitations in that it relies on an assumed Bayesian network as a generative model of protein function as well as on the available databases of protein functional annotations where term frequencies change over time. While the letter limitation can be remedied by more robust estimation of term frequencies in a large set of organisms, the performance accuracies in this setting are generally less comparable over two different CAFA experiments than in the precision-recall setting.

\section{Discussion}

Critical assessment challenges have been successfully adopted in a number of fields due to several factors. First, the recognition that improvements to methods are indeed necessary. Second, the ability of the community to mobilize enough of its members to engage in a challenge. Mobilizing a community is not a trivial task, as groups have their own research priorities and only a limited amount of resources to achieve them, which may deter them from undertaking a time-consuming and competitive effort a challenge may pose. At the same time, there are quite a few incentives to join a community challenge. Testing one's method objectively by a third party can establish credibility, help point out flaws, and suggest improvements. Engaging with other groups may lead to collaborations and other opportunities. Finally, the promise of doing well in a challenge can be a strong incentive heralding a group's excellence in their field. Since the assessment metrics are crucial to the performance of the teams, large efforts are made to create multiple metrics and to describe exactly what they measure. Good challenge organizers try to be attentive to the requests of the participants, and to have the rules of the challenge evolve based on the needs of the community. An understanding that a challenge's ultimate goal is to improve methodologies and that it takes several rounds of repeating the challenge to see results.

The first two CAFA challenges helped clarify that protein function prediction is a vibrant field, but also one of the most challenging tasks in computational biology. For example, CAFA provided evidence that the available function prediction algorithms outperform a straightforward use of sequence alignments in function transfer. The performance of methods in the Molecular Function category has consistently been reliable and also showed progress over time (unpublished results from CAFA2). On the other hand, the performance in the Biological Process or Cellular Component ontologies has not yet met expectations. One of the reasons for this may be that the terms in these ontologies are less predictable using amino acid sequence data and instead would rely more on high-quality systems data; e.g., see \cite{Costanzo2010}. The challenge has also helped clarify the problems of evaluation, both in terms of evaluating over consistent sub-graphs in the ontology but also in the presence of incomplete and biased molecular data. Finally, although it is still early, some best practices in the field are beginning to emerge. Exploiting multiple types of data is typically advantageous, although we have observed that both machine learning expertise and good biological insights tend to result in strong performance. Overall, while the methods in the Molecular Function ontology seem to be maturing, in part because of the strong signal in sequence data, the methods in the Biological Process and Cellular Component ontologies still appear to be in the early stages of development. With the help of better data over time, we expect significant improvements in these categories in the future CAFA experiments.

Overall, CAFA generated a strong positive response to the call for both challenge rounds, with the number of participants substantially growing between CAFA1 (102 participants) and CAFA2 (147). This indicates that there exists significant interest in developing computational protein function prediction methods, in understanding how well they perform, and in improving their performance. In CAFA2 we preserved the experiment rules, ontologies and metrics we used in CAFA1, but also added new ones to better capture the capabilities of different methods. The CAFA3 experiment will further improve evaluation by facilitating unbiased evaluation for several select functional terms.

More rounds of CAFA are needed to know if computational methods will improve as a direct result of this challenge. But given the community's growth and growing interest, we believe that CAFA is a welcome addition to the community of protein function annotators.

\section{Acknowledgements}
We thank Kymberleigh Pagel and Naihui Zhou for helpful discussions. This work was partially supported by NSF grants DBI-1458359 and DBI-1458477. 

\bibliographystyle{spmpsci}
\bibliography{cafa_chapter,refpedja}
\end{document}